\documentclass{aa}

\newcommand{\Msun}{\ensuremath{\,{\rm M}_\odot}}                  % Solar mass symbol
\newcommand{\Rsun}{\ensuremath{\,{\rm R}_\odot}}                  % Solar radius symbol
\newcommand{\psun}{\ensuremath{\,\rho_\odot}}                     % Solar density symbol
\newcommand{\Mjup}{\ensuremath{\,{\rm M}_{\rm Jup}}}              % Jupiter mass symbol
\newcommand{\Rjup}{\ensuremath{\,{\rm R}_{\rm Jup}}}              % Jupiter radius symbol
\newcommand{\pjup}{\ensuremath{\,\rho_{\rm Jup}}}                 % Jupiter density symbol
\newcommand{\Rearth}{\ensuremath{\,{\rm R}_\oplus}}               % Earth radius symbol
\newcommand{\Teff}{\ensuremath{T_{\rm eff}}}                      % Effective temperature symbol
\newcommand{\FeH}{\ensuremath{\textrm{[Fe/H]}}}                   % [Fe/H] symbol
\newcommand{\Porb}{\ensuremath{P_{\rm orb}}}                      % Orbital period symbol
\newcommand{\kms}{\,km\,s$^{-1}$}                                 % km/s symbol
\newcommand{\ms}{\,m\,s$^{-1}$}                                   % m/s^2 symbol
\newcommand{\mss}{\,m\,s$^{-2}$}                                  % m/s^2 symbol
\newcommand{\as}{\ensuremath{^{\prime\prime}}}                    % Arcsecond symbol
\newcommand{\er}[3]{\ensuremath{#1^{+#2}_{-#3}}}
\newcommand{\err}[5]{\ensuremath{{#1\,^{+#2}_{-#3}}\,^{+#4}_{-#5}}}
\newcommand{\reff}[1]{#1}

\usepackage{graphicx}
\usepackage{txfonts}
\usepackage{natbib}
\bibpunct{(}{)}{;}{a}{}{,}

\begin{document}
%%%%%%%%%%%%%%%%%%%%%%%%%%%%%%%%%%%%%%%%%%%%%%%%%%%%%%%%%%%%%%%%%%%%%%%%%%%%%%%%%%%%%%%%%%%%%%%%%%%%%%%%%%%%%%%%%%%%%%%%%%%%%%%%%%%%%%%%%%%%%%%%%%%%%

\title{A multiplicity study of transiting exoplanet host stars. II.}
\subtitle{Revised properties of transiting planetary systems with companions%
\thanks{Based on observations collected at the European Organisation for Astronomical Research in the Southern Hemisphere under ESO programmes 098.C-0589(A) and 099.C-0155(A).}}

\author{J.~Southworth\inst{1} \and A.~J.~Bohn\inst{2} \and M.~A.~Kenworthy\inst{2} \and C.~Ginski\inst{3} \and L.~Mancini\inst{4,5,6,7}}

\institute{Astrophysics Group, Keele University, Staffordshire ST5 5BG, UK  \\  \email{astro.js@keele.ac.uk}
           \and Leiden Observatory, Leiden University, PO Box 9513, 2300 RA Leiden, The Netherlands
           \and Sterrenkundig Instituut Anton Pannekoek, Science Park 904, 1098 XH Amsterdam, The Netherlands
           \and Department of Physics, University of Rome Tor Vergata, Via della Ricerca Scientifica 1, I-00133 Rome, Italy
           \and Max Planck Institute for Astronomy, K\"onigstuhl 17, D-69117 Heidelberg, Germany
           \and INAF -- Osservatorio Astrofisico di Torino, via Osservatorio 20, I-10025 Pino Torinese, Italy
           \and International Institute for Advanced Scientific Studies (IIASS), Via G.\ Pellegrino 19, I-84019 Vietri sul Mare (SA), Italy
       }

\date{Received \today\ / Accepted $<$date$>$}

\abstract
{Binarity is a widespread phenomenon around solar-type stars, including the host stars of transiting extrasolar planets.}
{We performed a detailed study of six transiting planetary systems with relatively bright stars close enough to affect observations of these systems. These contaminants were characterised in a companion work.}
{We used theoretical spectra to propagate the observed $K$-band light ratios into the optical passbands used to observe these systems. Light curves were analysed whilst taking the contaminating light and its uncertainty into account. We present and applied a method to correct the velocity amplitudes of the host stars for the presence of contaminating light.}
{We determined the physical properties of six systems (WASP-20, WASP-70, WASP-8, WASP-76, WASP-2, and WASP-131) whilst accounting for contaminating light. In the case of WASP-20, the measured physical properties are very different for the three scenarios considered: ignoring binarity, planet transits brighter star, and planet transits fainter star. In the other five cases, our results are very similar to those obtained when neglecting contaminating light. \reff{We used our results to determine the mean correction factors to planet radius, $\langle X_R \rangle$, mass, $\langle X_M \rangle$, and density, $\langle X_\rho \rangle$, caused by nearby objects. We find $\langle X_R \rangle = 1.009 \pm 0.045$, which is smaller than literature values because we were able to reject the possibility that the planet orbits the fainter star in all but one case. We find $\langle X_M \rangle = 1.031 \pm 0.019$, which is larger than $\langle X_R \rangle$ because of the strength of the effect of contaminating light on the radial velocity measurements of the host star. We find $\langle X_\rho \rangle = 0.995 \pm 0.046$: the small size of this correction is due to two effects: the corrections on planet radius and mass partially cancel; and some nearby stars are close enough to contaminate the light curves of the system but not radial velocities of the host star. These corrections can be applied to samples of transiting hot Jupiters to statistically remove biases due to light contamination.}}
{\reff{We conclude that} binarity of planet host stars is important for the small number of transiting \reff{hot Jupiters} with a very bright and close nearby star, but it has only a small effect on population-level studies of these objects.}


%%%%%%%%%%%%%%%%%%%%%%%%%%%%%%%%%%%%%%%%%%%%%%%%%%%%%%%%%%%%%%%%%%%%%%%%%%%%%%%%%%%%%%%%%%%%%%%%%%%%%%%%%%%%%%%%%%%%%%%%%%%%%%%%%%%%%%%%%%%%%%%%%%%%%
\maketitle
%%%%%%%%%%%%%%%%%%%%%%%%%%%%%%%%%%%%%%%%%%%%%%%%%%%%%%%%%%%%%%%%%%%%%%%%%%%%%%%%%%%%%%%%%%%%%%%%%%%%%%%%%%%%%%%%%%%%%%%%%%%%%%%%%%%%%%%%%%%%%%%%%%%%%

\section{Introduction}
\label{sec:intro}

The detection and characterisation of extrasolar planets is widespread and rapidly evolving. The vast majority of the early detections were via the radial velocity (RV) method, in which the orbital motion of the host star is observed \citep{MayorQueloz95nat,MarcyButler96apj,UdrySantos07araa}. This technique yields measurements of the orbital period, eccentricity, and separation, plus a lower limit on the mass of the planet. The dominant detection technique is currently the transit method, in which the drop in brightness of the host star due to the transit of the planet is observed. The transit method is useful for only a small fraction of planets, as the vast majority do not transit their host star, but it is highly efficient because thousands of stars can be surveyed simultaneously \citep[e.g.][]{Bakos+02pasp,Pollacco+06pasp,Borucki+10sci}. When combined for a single system, the RV and transit methods allow the full physical properties of the planetary system to be calculated: mass, radius, density, and surface gravity of both star and planet.

One of the basic assumptions of the transit method is that the light from the host star is not contaminated by light from a nearby star. If this assumption is incorrect, the measured physical properties of the system are biased away from their true values. The contamination causes a bias in two ways. Firstly, the contaminating light dilutes the light from the host star, thus decreasing the depth of the transit and leading to an underestimate of the radius of the transiting planet \citep[e.g.][]{Daemgen+09aa,Me10mn}. Secondly, spectral lines from the nearby star blend with those from the planet host star, which potentially cause the orbital motion of the star (and thus the mass of the planet) to be underestimated \citep{Buchhave+11apjs,Evans++16apj}. If ignored, contaminating light systematically affects the properties measured for populations of planets and their host stars, rendering unsafe any conclusions on the formation and evolution of planetary systems based on these demographics.

Our previous work on WASP-20 \citep{Evans++16apj} showed both effects very clearly, demonstrating the importance of correcting for contamination when determining the physical properties of a transiting planetary system. Using high-resolution adaptive-optics imaging, \citeauthor{Evans++16apj} showed that the WASP-20 system is composed of a resolved binary star, with a separation of $0.2578 \pm 0.0007$ arcsec, one of which is the host of a transiting planet \citep{Anderson+15aa}. Analysis of the available data yielded a mass and radius of the planet of $0.291 \pm 0.017$\Mjup\ and $1.20 \pm 0.14$\Rjup\ ignoring binarity, $0.378 \pm 0.022$\Mjup\ and $1.28 \pm 0.15$\Rjup\ if the planet orbits the brighter star, and $1.30 \pm 0.19$\Mjup\ and $1.69 \pm 0.12$\Rjup\ if the planet orbits the fainter star. This shows that binarity and light contamination can have a large effect on the measured properties of the planet.

\citet{Buchhave+11apjs} found comparable results for the planetary system Kepler-14, one component of a visual binary with a separation of 0.28\as\ and a magnitude difference of $\Delta V = 0.52 \pm 0.05$\,mag. These authors found that correcting for the presence of the nearby star increased the mass and radius of the planet by 60\% and 10\%, respectively. \citet{Buchhave+11apjs} were able to show that the planet orbits the brighter of the two stars by analysing the motion of the flux-weighted centroid of the binary during transit.

Another problem caused by contaminating light is the modification of the transmission spectrum of a transiting planet. A transmission spectrum is obtained by measuring the transit depth as a function of wavelength \citep{SeagerSasselov00apj,Brown01apj}. Unless the contaminating star has the same spectral energy distribution as the planet host star, its light could imprint a wavelength-dependent signal on the transit depth that could be erroneously interpreted as arising from the atmosphere of the planet. As an example, \citet{Me+15mn} found a strong Rayleigh scattering slope in the atmosphere of the planet WASP-103\,b. A faint nearby star was subsequently detected by \citet{WollertBrandner15aa}, and a reanalysis of the transit data by \citet{MeEvans16mn} yielded a significant modification to the transmission spectrum of the planet

Aside from the implications on measurements of the properties of planetary systems, the multiplicity of planet host stars is intrinsically interesting. Hot Jupiters cannot form in such tight orbits due to the high temperature and lack of mass available in this part of the protoplanetary disc \citep{Boss95sci,Lin++96nat} so must form further out and migrate inwards \citep[see][]{Baruteau+14prpl,Davies+14conf}. Smooth migration by interactions with the disc \citep{Lin++96nat} cannot explain the existence of hot Jupiters on eccentric or misaligned orbits \citep{WuMurray03apj,FabryckyTremaine07apj}. This suggests that gravitational interactions with a third body must be involved in the migration of at least a subset of hot Jupiters, either through planet-planet scattering events \citep{RasioFord96sci,Chatterjee+08apj} or the Kozai-Lidov mechanism \citep{FabryckyTremaine07apj,Naoz+11nat}. These predictions can be tested by assessing the fraction of planet host stars that are members of binary or higher-order multiple systems \citep{Knutson+14apj}.

Third bodies may also inhibit planet formation \citep{Fragner++11aa,Roell+12aa}. \reff{\citet{Kraus+16aj} found a paucity of binary companions to transiting planetary systems detected using the \textit{Kepler} satellite, compared to expectations from the binarity of field stars.} \citet{Ziegler+19aj} obtained high-resolution imaging of 524 planet candidates discovered using TESS \reff{(the Transiting Exoplanet Survey Satellite}; \citealt{Ricker+15jatis}) and found that the fraction of close companions was lower for projected separations $<$100\,au and higher for larger separations, to significance levels of 9.1$\sigma$ and 4.9$\sigma$ respectively. Similar trends were also found by \citet{Ngo+16apj} for 77 hot Jupiter systems. This implies that closer companions inhibit planet formation but that wider companions either aid planet formation and/or help the inward migration of planets to the relatively short orbital periods where they have been detected \reff{\citep[but see also][]{MoeKratter19xxx}}.

In a companion paper \citep[][hereafter Paper\,I]{Bohn+20aa} we presented high-resolution imaging observations of 45 transiting planet host stars, obtained using the SPHERE extreme adaptive-optics instrument on the Very Large Telescope (VLT) \citep{Beuzit+19aa}. We detected close companions in 26 systems, of which half were previously unknown. Our $K$-band contrast values were on average 7.0\,mag at 0.2\as\ and 8.9\,mag at separations beyond 1\as, allowing us to probe for companions down to the hydrogen-burning limit in the majority of our targets. The resulting multiplicity fraction of $55.4^{+5.9}_{-9.4}$\% is larger than but in agreement with previous assessments. In the current work we redetermined the physical properties of a subset of the targets from Paper\,I, accounting for the presence of the nearby companion. In several cases we analysed new photometry from space missions or from our own observations. Section\,\ref{sec:method} outlines our methods for photometric and spectroscopic observations, Section\,\ref{sec:res} presents our results, and Section\,\ref{sec:conc} summarises our work.

%%%%%%%%%%%%%%%%%%%%%%%%%%%%%%%%%%%%%%%%%%%%%%%%%%%%%%%%%%%%%%%%%%%%%%%%%%%%%%%%%%%%%%%%%%%%%%%%%%%%%%%%%%%%%%%%%%%%%%%%%%%%%%%%%%%%%%%%%%%%%%%%%%%%%

\section{Methods} \label{sec:method}

\begin{table*} \centering
\caption{\label{tab:sys} Summary of the high-resolution imaging results for the targets in this study.
Quantities are taken from Paper\,I. The objects are given in order of increasing $K$-band magnitude
difference ($\Delta K$), as this is the order in which they were analysed in the current work.}
\begin{tabular}{lccccc}
\hline
\hline
System & Primary star \Teff\ (K) & Separation (arcsec) & $\Delta K$ (mag) & Companion mass (\Msun) & Companion \Teff\ (K)  \\
\hline
WASP-20  & 6000 $\pm$ 100 & 0.259 $\pm$ 0.003 & 0.86 $\pm$ 0.06 & \er{0.89}{0.06}{0.07} & \er{5235}{242}{272} \\
WASP-70  & 5700 $\pm$  80 & 3.160 $\pm$ 0.004 & 1.38 $\pm$ 0.18 & \er{0.70}{0.06}{0.07} & \er{4504}{263}{213} \\
WASP-8   & 5600 $\pm$  80 & 4.520 $\pm$ 0.005 & 2.29 $\pm$ 0.08 & 0.53 $\pm$ 0.02       & \er{3758}{47}{43}   \\
WASP-76  & 6250 $\pm$ 100 & 0.436 $\pm$ 0.003 & 2.30 $\pm$ 0.05 & 0.78 $\pm$ 0.03       & \er{4824}{126}{128} \\
WASP-2   & 5170 $\pm$  60 & 0.710 $\pm$ 0.003 & 2.55 $\pm$ 0.07 & 0.40 $\pm$ 0.02       & \er{3523}{28}{19}   \\
WASP-131 & 5950 $\pm$ 100 & 0.189 $\pm$ 0.003 & 2.82 $\pm$ 0.20 & \er{0.62}{0.05}{0.04} & \er{4109}{200}{163} \\
\hline
\end{tabular}
\end{table*}

\subsection{Correcting the light curve for contamination} \label{sec:corr}

The amount of contaminating light is a standard parameter in the study of eclipsing binary systems, where it is typically referred to as `third light' \citep[e.g.][]{Kopal59book}. \reff{We use the definition that third light, $L_3$, is the fraction of the total light of the system arising from the third body, neglecting proximity effects in the inner binary system.} In order to include this in the model of the transit light curve, it is necessary to determine the amount of contaminating light in the passband used to obtain the light curve. This information is in general not directly measured, but can be determined by using synthetic spectra to extrapolate the flux ratio from the band it was measured in (in our case $K$) to the band the light curve was obtained in.

We interpolated within the grids of BT-Settl synthetic spectra to obtain spectra for the specific \Teff\ values of the presumed planet host star and the fainter nearby star. These were then scaled to the flux ratio we measured in the $K$-band, using the transmission profile for the SPHERE $K_s$ filter\footnote{\texttt{https://www.eso.org/sci/facilities/paranal/ instruments/sphere/inst/filters.html}}. Both spectra were then convolved with the profile of the relevant filter (see below for details) in order to determine the flux ratio in the passband used to obtain the transit light curve.

The uncertainties were propagated by repeating this analysis with the $K$-band magnitude difference perturbed by its upper and lower errorbar, and then applying this process to the \Teff s of both the planet host and the nearby star. The individual uncertainties were added in quadrature. The relevant quantities are reported in Table\,\ref{tab:sys}.

\subsubsection{Modelling the light curves}

We followed the precepts of the \textit{Homogeneous Studies} project \citep[see][and references therein]{Me12mn} to model the best available transit light curve for each target. We summarise the process here. The transits were fitted using the \textsc{jktebop} code \citep[see][and references therein]{Me13aa}, which parameterises the system using the fractional radii of the planet and its host star ($r_{\rm A} = \frac{R_{\rm A}}{a}$ and $r_{\rm b} = \frac{R_{\rm b}}{a}$ where $R_{\rm A}$ and $R_{\rm b}$ are the true radii of the star and planet and $a$ is the orbital semimajor axis), the orbital inclination ($i$), the orbital period (\Porb), and a reference time of mid-transit ($T_0$). Limb darkening was included using each of five parametric `laws' \citep[see][]{Me08mn} and third light was included using the values found in Section\,\ref{sec:corr}.

The fitted parameters in each case were the sum of the fractional radii, $r_{\rm A}+r_{\rm b}$, the ratio of the radii, $k = \frac{r_{\rm b}}{r_{\rm A}}$, $i$, and $T_0$. In some cases one limb darkening coefficient was also fitted. Their uncertainties were obtained using both Monte Carlo and residual permutation algorithms, and the larger of the two possibilities was adopted for each parameter. In all cases the uncertainty of the third light value was accounted for. The error estimates were increased to account for the variations between results obtained using the different limb darkening laws.

\subsection{Correcting the radial velocities for contamination} \label{sec:corr:rv}

The orbital motion of each planet host star has been measured as part of the process of confirming the planetary nature of the transiting companion. This was done by obtaining multiple high-resolution spectra, calculating the cross-correlation function (CCF) of each versus a numerical mask \citep{Baranne+96aas}, measuring the centroid of each CCF to obtain the RV, and fitting the RVs with a spectroscopic orbit. The amplitude of this orbit was then used in the measurement of the mass of the planet.

If the light from a nearby star contaminates the observed spectrum, it may bias the RVs measured from the CCFs away from the true value, affecting the measured mass of the planet. The size of this effect depends on multiple factors: (1) the light ratio of the contaminant versus the planet host star; (2) the fraction of light from the contaminant that enters the spectrograph slit or fibre; (3) the strength of the response of the spectrum of the contaminant to the numerical mask used to obtain the CCF; (4) the velocity difference between the host star and contaminant; and (5) the projected rotational velocities ($v\sin i$) of the two stars. The effect for point 2 is wavelength-dependent and thus is affected by both the spectral energy distribution of the two stars and the number of spectral lines involved in the RV measurement process as a function of wavelength.

This bias must be corrected for in order to measure the mass of the planet correctly, which means that it must be calculated. We constructed a simple model to estimated the correction factor for a system with a given light ratio, RV separation between the planet host and contaminant, and the $v\sin i$ values of the two stars. We used Gaussian functions to approximate the CCFs, with the expectation that this would induce significantly smaller inaccuracy than the assumptions we were forced to make on the spectral characteristics of the contaminating star (see above). More sophisticated simulations would ideally use true stellar spectra injected into the RV measurement pipeline for every observed spectrum, something outside the scope of the current work.

The correction factor was defined to be the true RV divided by the RV measured from the composite CCF. This definition means that the correction factors are usually above unity, and can become significantly larger than unity when the RV bias is large.

In each case, we used published measurements of the $v\sin i$ of the planet host star. In the absence of measurements of $v\sin i$ for the contaminating star we assumed a representative value of $2 \pm 1$\,km\,s$^{-1}$. The RV separation was taken to be the velocity amplitude of the host star's spectroscopic orbit, $K_{\rm A}$, which incurs two assumptions: the bias affecting the RVs away from quadrature scales linearly with RV separation; and the contaminating star is at the systemic velocity and thus is gravitationally bound to the planet host star on a wide orbit. We then generated Gaussian functions for the two CCFs, added them together, and fitted the composite CCF with a single Gaussian to determine the correction factor between the true and the measured RV of the planet host star.

The $v\sin i$ values used in this analysis were assumed to be full widths at half maximum (FWHMs). These were corrected to standard deviations, by dividing by $2\sqrt{2\ln2}$, in order to generate the Gaussian functions used for the CCFs.

Uncertainties in the correction factor were assessed by perturbing the input properties by their uncertainties (when known) or by a reasonable amount (when not known, and when possible). Linear interpolation was used in grids of correction factors in order to determine the value and uncertainty of the final number. A set of plots showing the behaviour of the correction factor is given during the discussion of WASP-20 below.

In all cases it must be borne in mind that the correction factor depends on our assumptions, and could even be negligibly different from unity if the RV of the contaminating star differs significantly from the systemic velocity of the planetary system. However, in the latter case, some systems would be picked up as having double-lined spectra indicative of either a contaminating star or an eclipsing binary system, so would be less likely to be ushered through the process of verifying that the transiting body is indeed a planet.

\subsection{Determining the physical properties of the systems} \label{sec:corr:absdim}

Once measured values of $r_{\rm A}$, $r_{\rm b}$, $i$, \Porb, and $K_{\rm A}$ were available, these were combined with tabulated predictions from theoretical evolution models for the properties of the planet host star \citep{Me09mn,Me10mn}. The physical properties of the systems were calculated using the velocity amplitude of the planet, $K_{\rm b}$, determined by iteratively maximising the agreement between the measured and predicted \Teff\ and $r_{\rm A}$ for the host star.

The uncertainties on the input parameters were propagated by a perturbation analysis to give statistical errorbars. The variations in results, from the use of five different sets of theoretical model predictions, were used to estimate the systematic errors for the output parameters. Both errorbars are given for all quantities that have a systematic error in their measurement.

%%%%%%%%%%%%%%%%%%%%%%%%%%%%%%%%%%%%%%%%%%%%%%%%%%%%%%%%%%%%%%%%%%%%%%%%%%%%%%%%%%%%%%%%%%%%%%%%%%%%%%%%%%%%%%%%%%%%%%%%%%%%%%%%%%%%%%%%%%%%%%%%%%%%%

\section{Results for individual systems} \label{sec:res}

The underestimation of the planet mass and radius is larger for stronger contamination, so in what follows we consider each planetary system in decreasing order of contamination level, until we reach those systems where the biases are negligible. It is important to remember that we cannot simply fit for the contamination level when modelling a transit light curve, as there is insufficient information in the light curve\footnote{It becomes possible to measure the contamination level when it contributes approximately 90\% of the total light or more; see \citet{Bognar+15aa}.} \citep[see][]{Me10mn}.

\subsection{WASP-20} \label{sec:lc:w20}

\begin{figure} \includegraphics[width=\columnwidth,angle=0]{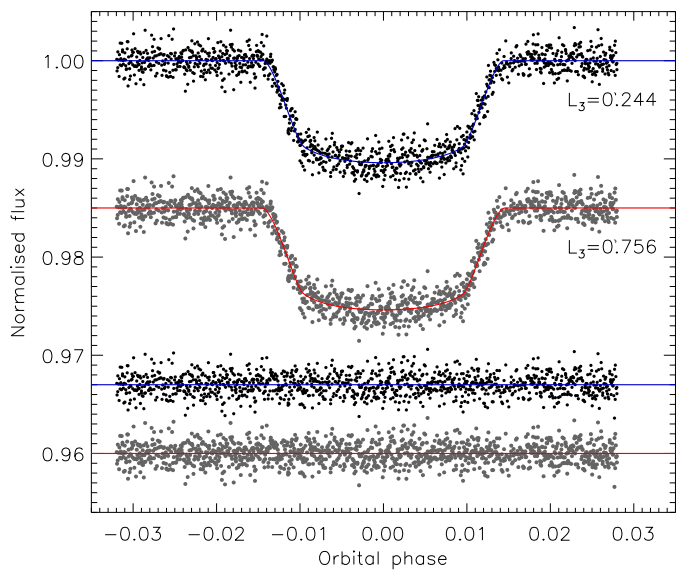}
\caption{\label{fig:lc:w20} Fits to the TESS light curve of WASP-20. The
observational data are shown as black and grey points. The {\sc jktebop} best
fit for the planet transiting star A is the blue line, and for transiting star
B is the red line. The residuals of the fits are shown at the base of the
figure with arbitrary offsets from zero.} \end{figure}

WASP-20 was previously presented as a poster child of the effect of contaminating light on the characterisation of a transiting planetary system \citep{Evans++16apj}. The discovery and first analysis of the system \citep{Anderson+15aa} proceeded under the assumption that the star was single, but an image from the SPHERE instrument showed it to be a double system separated by 0.26\as\ and with a magnitude difference of $\Delta K = 0.86$. \citet{Evans++16apj} modelled the best transit light curve then available for three scenarios: ignoring binarity; the planet orbits the brighter star; and the planet orbits the fainter star. The available data were insufficient to determine which of the last two scenarios was the correct one, although the planet-orbits-brighter-star was preferred. The measured mass and radius of the planet under these two scenarios differed by factors of 3.4 and 1.3, respectively. Both were also significantly different from the values obtained without accounting for the presence of contaminating light.

We have revisited this system for two reasons: a much better transit light curve is now available from the TESS satellite; and a more precise spectroscopic analysis of the host star has been published \citep{Andreasen+17aa}. The spectroscopic analysis was performed without accounting for contamination from the secondary star, so the results will be slightly biased; it is beyond the scope of the current work to account for this effect.

\citet{Andreasen+17aa} determined the \Teff\ of the WASP-20 system to be $5987 \pm 20$\,K. We took this to represent the brighter (and presumed planet host) star as it dominates the optical flux of the system. Using the $K$-band magnitude difference and contaminating star \Teff\ from Table\,\ref{tab:sys}, we determined a light ratio in the TESS passband of $0.323 \pm 0.063$. The contaminating light therefore contributes a fraction of $0.244 \pm 0.048$ of the light of the system.

The TESS data\footnote{Application: G011112, PI: J.\,Southworth} cover six transits in short cadence and were downloaded from the MAST archive\footnote{\texttt{https://mast.stsci.edu/portal/Mashup/Clients/Mast/ Portal.html}}. Each transit was extracted from the full light curve and normalised to unit flux by fitting a straight line to the adjacent out-of-transit data. The resulting data were modelled with the {\sc jktebop} code as described above and the results are given in Table\,\ref{tab:w20}.

\begin{figure} \includegraphics[width=\columnwidth,angle=0]{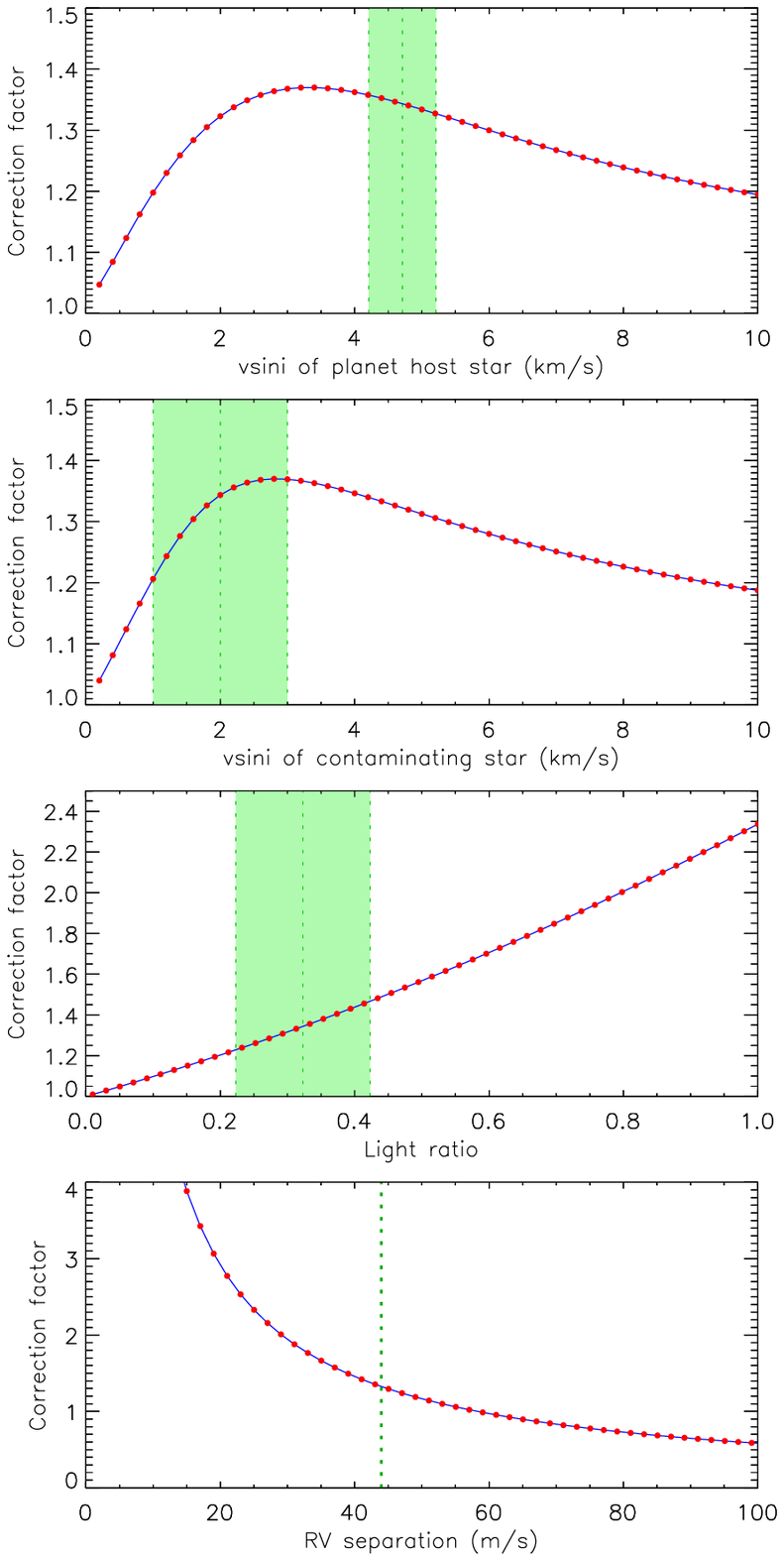}
\caption{\label{fig:lc:w20:cor} Plots showing the behaviour of the correction
factor for WASP-20 in the case that the planet transits the brighter star.
The correction factor is shown as a function of the $v\sin i$ values of the
two stars, and of the light ratio. The bottom panel is included for reference
and shows its variation as a function of the RV separation of the two stars.
In each case the correction factor is shown using red points connected by blue
lines, and the assumed system parameters are shown with their errorbars as
green dotted lines, with the range of values allowed by the uncertainties
indicated by light green shading.} \end{figure}

\begin{figure} \includegraphics[width=\columnwidth,angle=0]{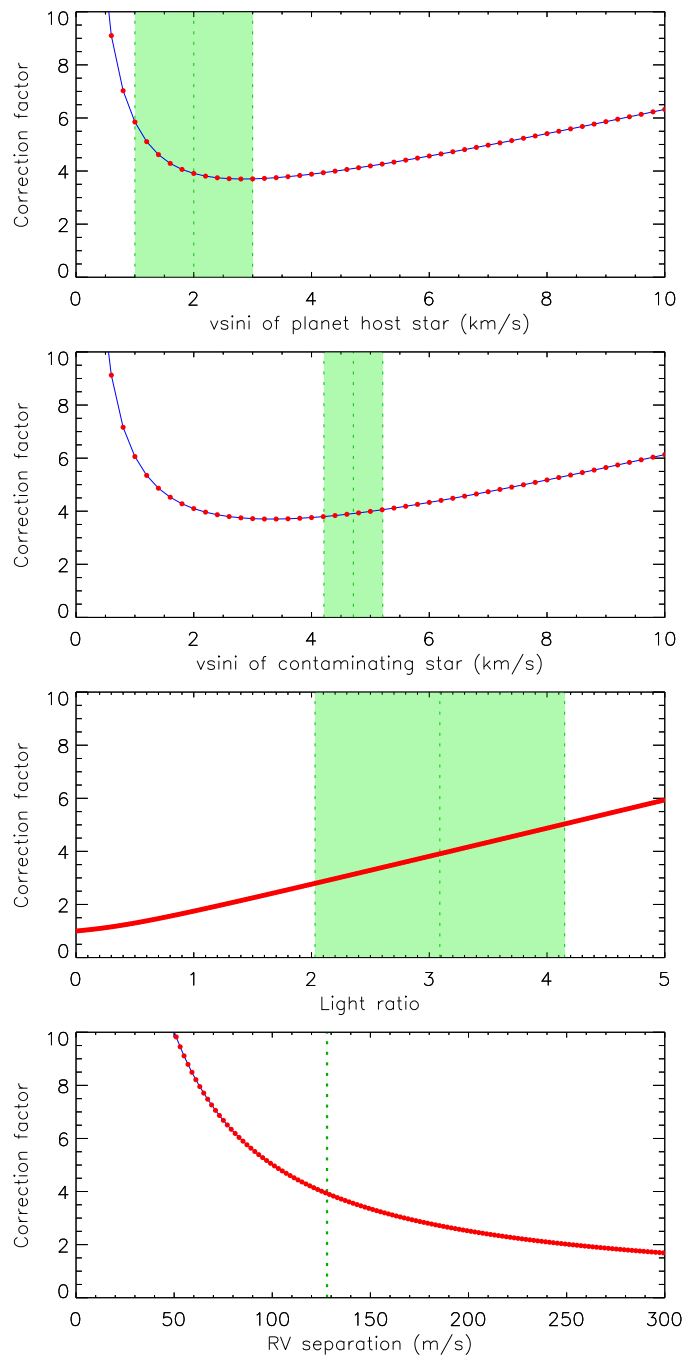}
\caption{\label{fig:lc:w20:corB} Plots showing the behaviour of the correction
factor for WASP-20 in the case that the planet host star is the fainter star.
Other comments are as for Fig.\,\ref{fig:lc:w20:cor}.} \end{figure}

\begin{table*} \centering
\caption{\label{tab:w20} Derived physical properties for the WASP-20 system. Where two sets of errorbars
are given, the first is the statistical uncertainty and the second is the systematic uncertainty.}
\begin{tabular}{llccc}
\hline
\hline
Parameter                       & Symbol                    & \citet{Anderson+15aa} & Planet transits brighter          & Planet transits                \\
                                &                           &                       & star (adopted solution)           & fainter star                   \\
\hline
Sum of the fractional radii     & $r_{\rm A}+r_{\rm b}$     &                       & $0.1100 \pm 0.0036$               & \er{0.0918}{0.0032}{0.0019}    \\
Ratio of the radii              & $k = R_{\rm b}/R_{\rm A}$ & $0.1079 \pm 0.0011$   & $0.1143 \pm 0.0040$               & $0.192 \pm 0.014$              \\
Inclination ($^\circ$)          & $i$                       & $85.56 \pm 0.22$      & $86.36 \pm 0.33$                  & $89.24 \pm 0.87$               \\
Fractional radius of the star   & $r_{\rm A} = R_{\rm A}/a$ & $0.1078 \pm 0.0027$   & $0.0987 \pm 0.0034$               & \er{0.0770}{0.0035}{0.0019}    \\
Fractional radius of the planet & $r_{\rm b} = R_{\rm b}/a$ &                       & $0.01129 \pm 0.00046$             & $0.01481 \pm 0.00080$          \\
\hline
Stellar mass (\Msun)            & $M_{\rm A}$               & $1.200 \pm 0.041$     & $1.113 \pm 0.027 \pm 0.021$       & $0.900 \pm 0.088$              \\
Stellar radius (\Rsun)          & $R_{\rm A}$               & $1.392 \pm 0.044$     & $1.242 \pm 0.044 \pm 0.008$       & $0.903 \pm 0.052$              \\
Stellar surface gravity (cgs)   & $\log g_{\rm A}$          & $4.231 \pm 0.020$     & $4.296 \pm 0.030 \pm 0.003$       & $4.481 \pm 0.043$              \\
Stellar density (\psun)         & $\rho_{\rm A}$            & $0.447 \pm 0.033$     & $0.581 \pm 0.060$                 & $1.22 \pm 0.17$                \\
Age (Gyr)                       & $\tau$                    & $7^{+2}_{-1}$         & \err{4.3}{0.8}{1.3}{0.9}{1.3}     & \err{3.3}{17.0}{0.2}{3.9}{2.1} \\
\hline
Planet mass (\Mjup)             & $M_{\rm b}$               & $0.311 \pm 0.017$     & $0.396 \pm 0.046 \pm 0.005$       & $0.998 \pm 0.087$              \\
Planet radius (\Rjup)           & $R_{\rm b}$               & $1.462 \pm 0.059$     & $1.382 \pm 0.057 \pm 0.008$       & $1.69 \pm 0.11$                \\
Planet surface gravity (\mss)   & $g_{\rm b}$               & $2.530 \pm 0.036$     & $5.13 \pm 0.73$                   & $8.7 \pm 1.1$                  \\
Planet density (\pjup)          & $\rho_{\rm b}$            & $0.099 \pm 0.012$     & $0.140 \pm 0.024 \pm 0.001$       & $0.193 \pm 0.034$              \\
Equilibrium temperature (K)     & $T_{\rm eq}$              & $1379 \pm 31$         & $1330 \pm 25$                     & $1027 \pm 52$                  \\
Orbital semimajor axis (au)     & $a$                       & $0.0600 \pm 0.0007$   & $0.05851 \pm 0.00047 \pm 0.00036$ & $0.0545 \pm 0.0018$            \\
\hline
\end{tabular}
\end{table*}

\subsubsection{Correction factor}

The star dominating the spectrum was found to have $v\sin i = 4.7 \pm 0.5$\,km\,s$^{-1}$ and a velocity amplitude $K_{\rm A} = 32.8 \pm 1.7$\,m\,s$^{-1}$ by \citet{Anderson+15aa}. We have assumed that this $v\sin i$ represents the brighter of the two stars: this assumption is questionable but a useful improvement would require effort beyond the scope of the current work. We used the light ratio of 0.323 but inflated the uncertainty for this value to 0.1 for the reasons described in Section\,\ref{sec:corr:rv}. We found a correction factor of 1.34 with errorbars of $\pm$0.02 from the $v\sin i$ of the brighter star, $\pm$0.08 from the $v\sin i$ of the fainter star, and $\pm$0.12 for the light ratio. Adding these uncertainties in quadrature gives a correction factor of $1.34 \pm 0.14$. This yields a velocity amplitude of $44.0 \pm 5.1$\,m\,s$^{-1}$, where the uncertainties from the measurement and from the correction factor have again been added in quadrature.

\citet{Evans++16apj} found a correction factor of $1.37 \pm 0.05$ for WASP-20 (after adjusting for their different definition of this quantity). This is in very good agreement with our value, despite the use of a different (actually more sophisticated) method and the erroneous use of $v\sin i = 2$\,km\,s$^{-1}$ for the brighter star. Our larger errorbar comes from the inclusion of more sources of uncertainty than those considered by \citet{Evans++16apj}, and it likely a better indicator of the intrinsic uncertainty of the correction factor.

If we turn to the alternative scenario of the planet orbiting the fainter star, we find a correction factor of $3.9 \pm 1.6$. The errorbar is the quadrature addition of individual errorbars of $\pm$1.1 from the $v\sin i$ of the fainter star, $\pm$0.1 from the $v\sin i$ of the brighter star, and $\pm$1.1 for the light ratio. \citet{Evans++16apj} found $5.56 \pm 0.63$ for this scenario, which is approximately 1$\sigma$ from our own value. The debiassed velocity amplitude from our correction factor is $128 \pm 7$\,m\,s$^{-1}$.

Fig.\,\ref{fig:lc:w20:cor} shows the results of an exploration of the dependence of the correction factor on the properties of the system assumed in its calculation, for the scenario where the brighter star hosts the planet. The top two panels show how it varies with the $v\sin i$ values of the host star and the contaminant. The dependence is relatively weak in the former case. We attribute this to the orbital motion being much smaller than the widths of the CCFs, so the precise positioning of the flux within the composite CCF does not have much effect on its measured centroid. The third panel shows that the correction factor is much more affected by the light ratio, as expected because a stronger dilution will naturally lead to a stronger bias in the measured RVs. The final panel is included for illustration, and shows how the correction factor depends on the RV separation of the system. The vertical dashed line indicates the true velocity amplitude of the planet host star as determined from the measured velocity amplitude and the correction factor found above. For reference, Fig.\,\ref{fig:lc:w20:corB} shows the variation of the correction factor in the case that the planet orbits the fainter of the two stars.

\subsubsection{Physical properties} \label{sec:lc:w20:absdim}

We determined the physical properties of the WASP-20 planetary system under both scenarios: planet orbits brighter star and planet orbit fainter star. \citet{Andreasen+17aa} determined the \Teff\ of the WASP-20 system to be $5987 \pm 20$\,K, and we used this value in preference to the value of $6000 \pm 100$\,K adopted for Paper\,I. We inflated the errorbar to 50\,K as this is the level of variation between different high-quality analyses of similar stars \citep[e.g.][]{Depascale+14aa,Gomez+14aa,Ryabchikova+16mn}. \citeauthor{Andreasen+17aa} also quoted a metallicity of $\FeH = 0.07 \pm 0.02$; we have adopted this with a larger errorbar of 0.05\,dex for similar reasons \citep[see e.g.][]{Jofre+14aa,Depascale+14aa}.

Physical properties were obtained for the two scenarios, using the method outline in Section\,\ref{sec:corr:absdim}. We used the respective \Teff\ values of the two stars and, under the assumption of physical relation, the same metallicity value. Table\,\ref{tab:w20} contains the results and shows that the measured planet properties change significantly between the two scenarios. The mass of the planet is most affected, being $0.40 \pm 0.05$\Mjup\ if the planet orbits the brighter star and $1.00 \pm 0.09$\Mjup\ if it orbits the fainter. This is in good agreement with the results of \citet{Evans++16apj}. The inclusion of the TESS data in our analysis has allowed the radius of the planet to be measured to a greater precision compared to previous work.

\subsection{WASP-70} \label{sec:lc:w70}

\begin{figure} \includegraphics[width=\columnwidth,angle=0]{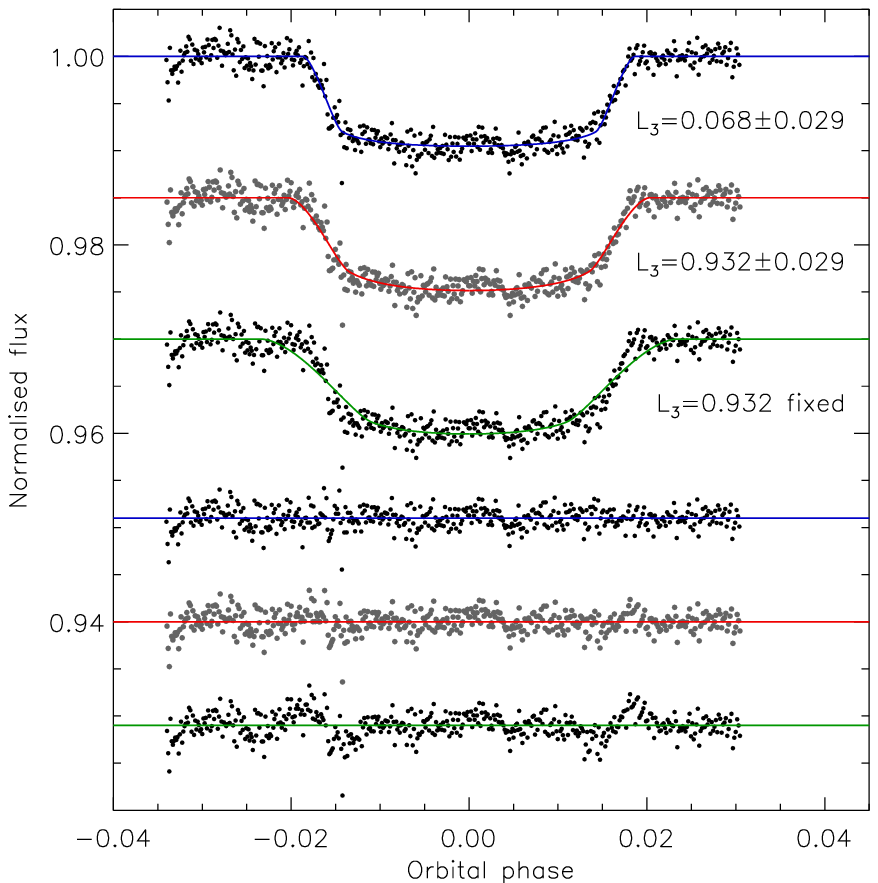}
\caption{\label{fig:lc:w70} Fit to the Euler telescope light curve of WASP-70. The observational
data are shown as black and grey points. \reff{The {\sc jktebop} fits are shown for three scenarios:
planet transiting brighter star; planet transiting fainter star; and planet transiting fainter star
with the third light value forced to match that found from our direct image. The residuals of the
fits are shown at the base of the figure with arbitrary offsets from zero.}} \end{figure}

WASP-70 is the system with the second-brightest nearby companion, with $\Delta K = 1.38 \pm 0.18$\,mag and a separation of 3.2\as. The companion was detected in the discovery paper of the system \citep{Anderson+14mn} and in subsequent Lucky Imaging \citep{WollertBrandner15aa,Ginski+16mn,Evans+18aa} and adaptive-optics \citep{Ngo+16apj} studies. In Paper\,I we established that its proper motion is consistent with it being a bound, not background, object. \citet{Anderson+14mn} accounted for the companion star by removing its contribution from the light curves, thus the uncertainty in its measured contribution was ignored.

WASP-70 has not been observed using TESS due to its equatorial sky position, so we have analysed the system based on the best light curve we are aware of: the $r$-band EulerCam data from the night of 2011 September 20 \citep{Anderson+14mn}. We determined an $r$-band light ratio of $0.073 \pm 0.032$ using the $i$-band magnitude difference of $\Delta i = 2.62 \pm 0.18$ from \citet{WollertBrandner15aa} -- this is preferable to our own $\Delta K$ value as it has the same precision and is much closer in wavelength.

The data were modelled and the system parameters were determined in the same way as for WASP-20. The results of this process are given in Table\,\ref{tab:w70}. We found that the limb darkening predicted by theoretical studies is too strong for this light curve, so we fitted for the linear coefficient in our final analyses.

Light curve fits for the scenario where the planet transits the fainter star are strongly disfavoured: the best fit we found had $\chi^2 = 540$, versus $\chi^2 = 467$ under the assumption that the planet transits the brighter star. Both $\chi^2$ values are for 467 datapoints and six fitted parameters, after the errorbars of the datapoints had been rescaled to give a reduced $\chi^2$ of 1.0 for the best fits. We therefore rule out the possibility that the planet transits the fainter companion star. This supports our previous findings that third light is perfectly degenerate with other parameters of the fit \citep{Me11mn} except in cases where it is at least 90\% of the total light of the system \citep{Bognar+15aa}.

\begin{table} \centering
\caption{\label{tab:w70} Derived physical properties for the WASP-70 system. Where two sets of errorbars
are given, the first is the statistical uncertainty and the second is the systematic uncertainty.}
\begin{tabular}{lccc}
\hline
\hline
Parameter                       & \citet{Anderson+14mn}     & This work (planet              \\       % Planet transits star A & Planet transits star B
                                &                           & transits brighter star)        \\       % (adopted solution)     &
\hline                                                                                                %
$r_{\rm A}+r_{\rm b}$           &                           & $0.1312 \pm 0.0083$            \\       % & $0.1243^{+0.0041}_{-0.0027}$
$k$                             & $0.0985 \pm 0.0012$       & $0.0976 \pm 0.0020$            \\       % & $0.206^{+0.031}_{-0.019}$
$i$ ($^\circ$)                  & $87.12^{+1.24}_{-0.65}$   & $86.5 \pm 0.9$                 \\       % & $89.6 \pm 1.0$
$r_{\rm A}$                     &                           & $0.1196 \pm 0.0075$            \\       % & $0.1031^{+0.0041}_{-0.0021}$
$r_{\rm b}$                     &                           & $0.01166 \pm 0.00086$          \\       % & $0.0213^{+0.0030}_{-0.0020}$
\hline                                                                                                %
$M_{\rm A}$             (\Msun) & $1.106 \pm 0.042$         & $1.111 \pm 0.029 \pm 0.017$    \\       % &
$R_{\rm A}$             (\Rsun) & $1.215^{+0.064}_{-0.069}$ & $1.251 \pm 0.079 \pm 0.006$    \\       % &
$\log g_{\rm A}$        (cgs)   & $4.314^{+0.052}_{-0.036}$ & $4.290 \pm 0.055 \pm 0.002$    \\       % &
$\rho_{\rm A}$          (\psun) & $0.619^{+0.136}_{-0.077}$ & $0.57 \pm 0.11$                \\       % &
$\tau$                  (Gyr)   &                           & \err{4.4}{0.7}{1.3}{1.2}{1.6}  \\       % &
\hline                                                                                                %
$M_{\rm b}$             (\Mjup) & $0.590 \pm 0.022$         & $0.592 \pm 0.019 \pm 0.006$    \\       % &
$R_{\rm b}$             (\Rjup) & $1.164^{+0.073}_{-0.102}$ & $1.186 \pm 0.088 \pm 0.006$    \\       % &
$g_{\rm b}$             (\mss)  & $10.0^{+1.4}_{-1.1}$      & $10.4 \pm 1.6$                 \\       % &
$\rho_{\rm b}$          (\pjup) & $0.375^{+0.104}_{-0.060}$ & $0.332 \pm 0.076 \pm 0.002$    \\       % &
$T_{\rm eq}$            (K)     & $1376 \pm 40$             & $1433 \pm 46$                  \\       % &
$a$                     (au)    & $0.04853 \pm 0.00062$     & $0.0486 \pm 0.0004 \pm 0.0003$ \\       % &
\hline
\end{tabular}
\end{table}

\subsubsection{Physical properties}

The angular separation of the planet host star and the contaminant, 3.2\as, is significantly larger than the diameter of the optical fibres used to feed the CORALIE and HARPS spectrographs (2.0\as\ and 1.0\as, respectively). We have therefore assumed that the brighter star is the planet host, and that there is no need to correct the RVs of the system for the presence of the fainter star.

\citet{Sousa+18aa} recently presented new atmospheric parameters for the host star: $\Teff = 5864 \pm 25$\,K and $\FeH = 0.21 \pm 0.02$. As in Section\,\ref{sec:lc:w20:absdim} we have adopted larger errorbars of 50\,K and 0.05\,dex for these quantities, respectively. Our results are given in Table\,\ref{tab:w70} and are in good agreement with those of \citet{Anderson+14mn}. This is unsurprising, as \citeauthor{Anderson+14mn} accounted for the presence of the companion in their work. It is also encouraging, because \citeauthor{Anderson+14mn} did not account for the uncertainty in the light contributed by the companion and thus potentially neglected an important source of uncertainty. We do find the star to be slightly larger: this causes an increase in the measured radius and equilibrium temperature of the planet, and a decrease in its measured density.

\subsection{WASP-8} \label{sec:lc:w08}

\begin{figure} \includegraphics[width=\columnwidth,angle=0]{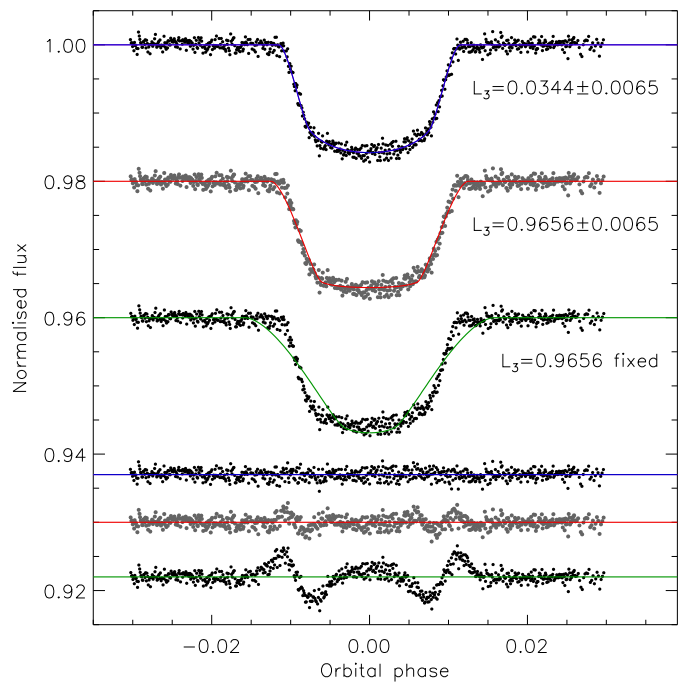}
\caption{\label{fig:lc:w08} Fits to the TESS light curve of WASP-8. The
observational data are shown as black and grey points. \reff{The plotted
quantities are otherwise the same as for Fig.\,\ref{fig:lc:w70}.}} \end{figure}

WASP-8 has the third-brightest nearby companion, with $\Delta K = 2.29 \pm 0.08$\,mag and a separation of 4.52\as. The companion was detected in the discovery paper of the system \citep{Queloz+10aa}, who did not comment on how (or whether) its presence was accounted for in their analysis. It was also detected in a subsequent adaptive-optics study \citep{Ngo+15apj} and a Lucky-Imaging study \citep{Evans+16aa}, and is visible in 2MASS images \citep{Queloz+10aa}. The Gaia DR2 database \citep{Gaia18aa} lists parallaxes and proper motions of the two objects that are consistent with each other, supporting their companionship.

WASP-8 has been observed using TESS and this light curve was treated in the same way as the one for WASP-20 (Section\,\ref{sec:lc:w20}). Our $\Delta K$ value corresponds to a light ratio between the two stars of $0.0356 \pm 0.0067$ and thus a third light of $L_3 = 0.0344 \pm 0.00065$. This was used to model the TESS light curve under the two alternative possibilities of which is the host star. WASP-8 has an eccentric orbit so we constrained the Poincar\'e elements to be $e\cos\omega = 0.02307 \pm 0.00010$ and $e\sin\omega = -0.0392 \pm 0.0029$ \citep{Queloz+10aa}.

As with WASP-70, we find that light curve fits for the scenario where the planet transits the fainter star are strongly disfavoured, with $\chi^2 = 1090$ versus $\chi^2 = 674$ for 668 degrees of freedom. These numbers again refer to the case where the errorbars of the TESS data, which are far too small, were rescaled to give a reduced $\chi^2$ of approximately 1.0. There is also a large tension in the planet-transits-fainter-star scenario as the best-fitting value of third light disagrees with the applied prior at the 10$\sigma$ level. The best fits for the two options (Fig.\,\ref{fig:lc:w08}) clearly show a poor fit to the large-$L_3$ scenario. We therefore proceeded under the safe assumption that the planet transits the brighter star.

\begin{table} \centering
\caption{\label{tab:w08} Derived physical properties for the WASP-8 system. Where two sets of errorbars
are given, the first is the statistical uncertainty and the second is the systematic uncertainty.}
\begin{tabular}{lcc}
\hline
\hline
Parameter                 & \citet{Queloz+10aa}             & This work (planet              \\
                          &                                 & transits brighter star)        \\
\hline
$r_{\rm A}+r_{\rm b}$     &                                 & $0.0623 \pm 0.0013$            \\
$k$                       & $0.1130^{+0.0015}_{-0.0013}$    & $0.1227 \pm 0.0011$            \\
$i$ ($^\circ$)            & $88.55^{+0.15}_{-0.17}$         & $88.51 \pm 0.09$               \\
$r_{\rm A}$               & $0.0549 \pm 0.0024$             & $0.0555 \pm 0.0011$            \\
$r_{\rm b}$               & $0.00620^{+0.00036}_{-0.00033}$ & $0.00681 \pm 0.00018$          \\
\hline
$M_{\rm A}$       (\Msun) & $1.030^{+0.054}_{-0.061}$       & $1.093 \pm 0.024 \pm 0.023$    \\
$R_{\rm A}$       (\Rsun) & $0.945^{+0.051}_{-0.036}$       & $0.976 \pm 0.020 \pm 0.007$    \\
$\log g_{\rm A}$  (cgs)   & $4.5 \pm 0.1$                   & $4.498 \pm 0.018 \pm 0.003$    \\
$\rho_{\rm A}$    (\psun) & $1.22^{+0.17}_{-0.15}$          & $1.176 \pm 0.070$              \\
$\tau$            (Gyr)   & 3--5                            & \err{0.3}{0.9}{0.0}{0.1}{0.1}  \\
\hline
$M_{\rm b}$       (\Mjup) & $2.244^{+0.079}_{-0.093}$       & $2.216 \pm 0.035 \pm 0.031$    \\
$R_{\rm b}$       (\Rjup) & $1.038^{+0.007}_{-0.047}$       & $1.165 \pm 0.032 \pm 0.008$    \\
$g_{\rm b}$       (\mss)  &                                 & $42.5 \pm  2.3$                \\
$\rho_{\rm b}$    (\pjup) &                                 & $1.31 \pm 0.10 \pm 0.01$       \\
$T_{\rm eq}$      (K)     &                                 & $947 \pm 12$                   \\
$a$               (au)    & $0.0801^{+0.0014}_{-0.0016}$    & $0.0817 \pm 0.0006 \pm 0.0006$ \\
\hline
\end{tabular}
\end{table}

\subsubsection{Physical properties}

The angular separation of the planet host star and the contaminant is 4.5\as\ so, like WASP-70, is significantly larger than the entrance apertures of the spectrographs. We have therefore not corrected $K_{\rm A}$ for the presence of the nearby star. \citet{Mortier+13aa} gave the atmospheric parameters as $\Teff = 5690 \pm 36$\,K and $\FeH = 0.29 \pm 0.03$; we have adopted larger errorbars as in Section\,\ref{sec:lc:w20:absdim}.

Our physical properties were calculated with these atmospheric parameters, the value of $K_{\rm A} = 221.1 \pm 1.2$\ms\ given by \citet{Knutson+14apj}, and the photometric parameters we determined from the TESS data above (Table\,\ref{tab:w08}). These give a significant improvement in the precision of the measured properties compared to those quoted by \citet{Queloz+10aa}, primarily due to the availability of the TESS data. However, the high density of the star implies a rather young age and this puts it near the edge of the grids of theoretical stellar evolutionary models used in our study. This in turn causes a larger systematic error in the physical properties compared to the other stars in the current work. The young age is also in poor agreement with the lithium abundance determined by \citet{Queloz+10aa}, a discrepancy which should be investigated in future using other age indicators such as kinematic properties and emission in the calcium H and K lines.

\subsection{WASP-76} \label{sec:lc:w76}

\begin{figure} \includegraphics[width=\columnwidth,angle=0]{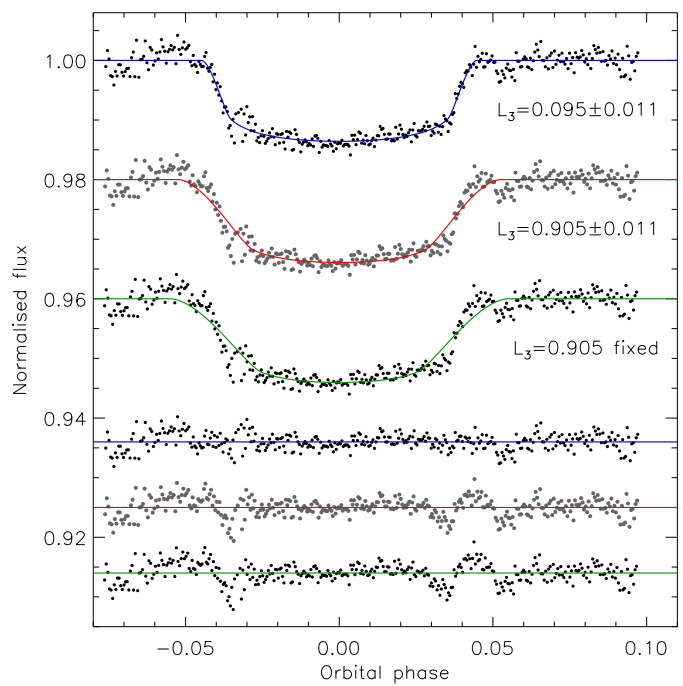}
\caption{\label{fig:lc:w76} Fit to the light curve of WASP-76 presented in the current work.
The plotted quantities are otherwise the same as for \reff{Fig.\,\ref{fig:lc:w70}}.} \end{figure}

The companion of WASP-76 has a very similar magnitude difference to that of WASP-8, $\Delta K = 2.30 \pm 0.05$\,mag, but a much smaller separation of 0.436\as. The companion was not detected in the discovery paper of the system \citep{West+16aa}, but was found in a later work \citep{WollertBrandner15aa}. It has been redetected in subsequent studies \citep{Ginski+16mn,Ngo+16apj} and in Paper\,I we confirmed the common proper motion of the two objects.

In light of this it is worthwhile to reconsider the properties of WASP-76. However, it has not been observed using TESS and the available light curves \citep{West+16aa} are either incomplete or riven with red noise. We therefore obtained a new transit light curve of WASP-76, and used this to refine the properties of the system. The light curve was observed on the night of 2017 October 26 using the CAHA 1.23\,m telescope. The data were obtained with the telescope defocussed to increase the photometric precision \citep[see][]{Me+09mn}, through a Cousins $R$ filter, and using the standard approach of our group \citep[e.g.][]{Ciceri+15aa2,Mancini+17mn}. Data reduction was performed using the {\sc defot} pipeline \citep{Me+09mn,Me+14mn}, yielding differential magnitudes relative to an optimal ensemble of comparison stars with timestamps on the BJD(TDB) timescale.

% Light ratio:       0.10460283     0.012265435
% Light contribution:      0.094697234     0.011103933

Our $\Delta K$ value corresponds to a light ratio between the two stars of $0.105 \pm 0.012$ and thus a third light of $L_3 = 0.095 \pm 0.011$. This was used to model our $R$-band light curve under the two alternative possibilities of which is the host star. As with WASP-8 (Sect.\,\ref{sec:lc:w08}), the fits for the planet-transits-fainter-star scenario are heavily disfavoured and can be discounted (see Fig.\,\ref{fig:lc:w76}).

As an additional result, the transit we observed appeared approximately 8.6 minutes earlier than predicted by the ephemeris from \citet{West+16aa}. We therefore provide a revised orbital ephemeris for this system:
% $$ T_0 = {\rm BJD(TDB)} \,\, 2\,458\,053.47655 (8) \, + \, 1.8095695 (3) E $$
$$ T_0 = {\rm BJD(TDB)} \,\, 2\,458\,053.47655 (34) \, + \, 1.8098798 (5) E $$
where $E$ is the cycle number since the reference time and the bracketed quantities give the uncertainty in the last digit of the previous number. This ephemeris is based on our own transit and the reference time of mid-transit quoted by \citet{West+16aa}. The orbital period is 0.54\,s shorter than the one given in \citet{West+16aa}, a change significantly larger than the errorbars. Further times of transit should be obtained in order to check if this system shows deviations from a constant orbital period. At present, the hypothesis that there are transit timing variations in the system is completely degenerate with the hypothesis that the errorbars for the original ephemeris are underestimated. The law of parsimony indicates we should assume the latter option at present.

% West et al (2016A+A...585A.126W)
% P (d) 1.809886 ± 0.000001
% Tc (HJD) (UTC) 2456107.85507 ± 0.00034
% Numbers are the same as in the preprint (xxx13105607)

% Our CAHA data:
%  20  Ephemeris timebase     58053.4765483004 +/-   0.0003535733

%     print, (1.809886d0-1.8098798d0)/0.000001d0, (1.809886d0-1.8098798d0)*86400.0d0
%       6.2000000      0.53568000

% period.pro output:  58053.476550 +/- 0.000080   1.809879750 +/- 0.000000325

\begin{table} \centering
\caption{\label{tab:w76} Derived physical properties for the WASP-76 system. Where two sets of errorbars
are given, the first is the statistical uncertainty and the second is the systematic uncertainty.}
\begin{tabular}{lcc}
\hline
\hline
Parameter                 & \citet{West+16aa}               & This work (planet               \\
                          &                                 & transits brighter star)         \\
\hline
$r_{\rm A}+r_{\rm b}$     &                                 & $0.276^{+0.014}_{-0.004}$       \\
$k$                       & $0.1090 \pm 0.0007$             & $0.1126^{+0.0045}_{-0.0022}$    \\
$i$ ($^\circ$)            & $88.0^{+1.3}_{-1.6}$            & $89.9^{+0.1}_{-4.3}$            \\
$r_{\rm A}$               &                                 & $0.248^{+0.012}_{-0.004}$       \\
$r_{\rm b}$               &                                 & $0.0280^{+0.0017}_{-0.0006}$    \\
\hline
$M_{\rm A}$       (\Msun) & $1.46 \pm 0.07$                 & \err{1.356}{0.048}{0.025}{0.009}{0.014}           \\
$R_{\rm A}$       (\Rsun) & $1.73 \pm 0.04$                 & \err{1.716}{0.086}{0.030}{0.004}{0.006}           \\
$\log g_{\rm A}$  (cgs)   & $4.128 \pm 0.015$               & \err{4.101}{0.015}{0.041}{0.001}{0.001}           \\
$\rho_{\rm A}$    (\psun) & $0.186^{+0.008}_{-0.018}$       & \er{0.268}{0.013}{0.035}                          \\
$\tau$            (Gyr)   &                                 & \err{1.0}{0.3}{0.8}{0.2}{0.2}                     \\
\hline
$M_{\rm b}$       (\Mjup) & $0.92 \pm 0.03$                 & \err{0.914}{0.025}{0.017}{0.004}{0.006}           \\
$R_{\rm b}$       (\Rjup) & $1.83^{+0.06}_{-0.04}$          & \err{1.885}{0.117}{0.042}{0.004}{0.006}           \\
$g_{\rm b}$       (\mss)  & $6.31 \pm 0.39$                 & \er{6.38}{0.30}{0.72}                             \\
$\rho_{\rm b}$    (\pjup) & $0.151 \pm 0.010$               & \err{0.1276}{0.0088}{0.0208}{0.0004}{0.0003}      \\
$T_{\rm eq}$      (K)     & $2160 \pm 40$                   & \er{2235}{  56}{  25}                             \\
$a$               (au)    & $0.0330 \pm 0.0005$             & \err{0.03217}{0.00038}{0.00020}{0.00007}{0.00011} \\
\hline
\end{tabular}
\end{table}

\subsubsection{Physical properties}

The angular separation of the planet host star and the contaminant is 0.44\as\ so we operated under the assumption that the fainter star fully contaminated the spectrum and therefore the value of $K_{\rm A}$ must be corrected for this. We adopted $K_{\rm A} = 112 \pm 1$\ms\ and $v \sin i = 2.33 \pm 0.36$\kms\ from \citet{Brown+17mn} as the best estimates of these quantities, as they are based on a sophisticated modelling of the RVs and spectral line deformation during transit.

We found a correction factor of 1.111 with errorbars of $\pm$0.006 from the $v\sin i$ of the brighter star, $\pm$0.007 from the $v\sin i$ of the fainter star, and $\pm$0.013 for the light ratio. Adding these uncertainties in quadrature gives a correction factor of $1.111 \pm 0.016$. This yields a velocity amplitude of $124.4 \pm 1.8$\,m\,s$^{-1}$, where the uncertainties from the measurement and from the correction factor have again been added in quadrature. This change in $K_{\rm A}$ is modest but nevertheless significant at the 7$\sigma$ level.

\citet{Andreasen+17aa} gave atmospheric properties for the planet host star of $\Teff = 6347 \pm 52$\,K and $\FeH = 0.36 \pm 0.04$; we adopted a errorbar of 0.05\,dex for \FeH\ (see Section\,\ref{sec:lc:w20:absdim}). With these properties, the corrected $K_{\rm A}$, and the photometric parameters from our new light curve, we determined the physical properties of the system and give these in Table\,\ref{tab:w76}. The new light curve gives a lower density and thus lower mass for the star, which balances the corrected value of $K_{\rm A}$ so the measured planet mass is almost unchanged. However, the smaller semimajor axis and higher \Teff\ of the host star adopted in the current study yields a significantly hotter equilibrium temperature of the planet of \er{2235}{56}{25}\,K.

WASP-76\,b is one of the hottest planets known. As a result of this, its atmosphere has been the subject of several observational studies \citep{Tsiaras+18aj,Seidel+19aa,Zak+19aj}. These should be revisited now that the planetary system is known to have a close companion (Paper\,I) and the planet itself has a larger radius and higher measured equilibrium temperature (this work).

\subsection{WASP-2} \label{sec:lc:w2}

\begin{table} \centering
\caption{\label{tab:w02} Derived physical properties for the WASP-2 system. Where two sets of errorbars
are given, the first is the statistical uncertainty and the second is the systematic uncertainty.}
\setlength{\tabcolsep}{4pt}
\begin{tabular}{lcc}
\hline
\hline
Parameter                 & Value                      & Reference \\
\hline
$r_{\rm A}+r_{\rm b}$     & $0.1403 \pm 0.0021$            & \citet{Me+10mn} \\
$k$                       & $0.1326 \pm 0.0007$            & \citet{Me+10mn} \\
$i$ ($^\circ$)            & $84.81 \pm 0.17$               & \citet{Me+10mn} \\
$r_{\rm A}$               & $0.1238 \pm 0.0018$            & \citet{Me+10mn} \\
$r_{\rm b}$               & $0.01643 \pm 0.00030$          & \citet{Me+10mn} \\
\hline
$M_{\rm A}$       (\Msun) & $0.843 \pm 0.033 \pm 0.019$    & This work \\
$R_{\rm A}$       (\Rsun) & $0.821 \pm 0.013 \pm 0.006$    & This work \\
$\log g_{\rm A}$  (cgs)   & $4.536 \pm 0.015 \pm 0.003$    & This work \\
$\rho_{\rm A}$    (\psun) & $1.524 \pm 0.067$              & This work \\
$\tau$            (Gyr)   & \err{7.6}{2.5}{3.3}{3.2}{4.1}  & This work \\
\hline
$M_{\rm b}$       (\Mjup) & $0.892 \pm 0.027 \pm 0.013$    & This work \\
$R_{\rm b}$       (\Rjup) & $1.060 \pm 0.024 \pm 0.008$    & This work \\
$g_{\rm b}$       (\mss)  & $19.70 \pm  0.78$              & This work \\
$\rho_{\rm b}$    (\pjup) & $0.701 \pm 0.041 \pm 0.005$    & This work \\
$T_{\rm eq}$      (K)     & $1286 \pm   17$                & This work \\
$a$               (au)    & $0.0308 \pm 0.0004 \pm 0.0002$ & This work \\
\hline
\end{tabular}
\end{table}

WASP-2 has a faint companion at an angular separation of 0.710\as\ that was discovered in the process of confirming the planetary nature of the system \citep{Cameron+07mn}. The companion has been detected by several follow-up surveys \citep{Daemgen+09aa,Bergfors+13mn,Ngo+15apj,WollertBrandner15aa}. \citet{Evans+16aa} confirmed that the two objects have the same proper motion to 5$\sigma$ significance, and tentatively identified orbital motion.

WASP-2 has not so far been observed using the TESS satellite, so the best transit light curves available are observations of three transits using the telescope-defocussing method by \citet{Me+10mn}. These authors accounted for the presence of the nearby star in their analysis, with a magnitude difference very similar to that found in Paper\,I. There is no need to repeat this work, so we use their values of $r_1$, $r_2$ and $i$ in our analysis.

However, the RVs of WASP-2\,A have not been corrected for the effects of contamination from the nearby star, and this is why we have revisited the system here. The separation of the two stars is significantly smaller than that angular size of the optical fibre used to obtain the RVs by \citet{Cameron+07mn}, and comparable to the slit width for the observations presented by \citet{Knutson+14apj}, which were obtained using the Keck telescope and HIRES spectrograph. We have assumed that the fainter star fully contaminates the spectrum of the planet host star in order to calculate the correction factor to $K_{\rm A}$: if the contamination is smaller then the correction factor would be decreased approximately linearly so it would be easy to adjust these results in future.

To calculate the correction factor we adopted $v\sin i = 1.3 \pm 0.5$\kms\ from the Rossiter-McLaughlin analysis of \citet{Albrecht+11apj2}, $K_{\rm A} = 156.7 \pm 1.2$\ms\ from \citet{Knutson+14apj}, and $\Teff = 5170 \pm 60$\,K from \citet{Me12mn}. The light ratio of the system, determined as in Section\,\ref{sec:corr}, is $0.0204 \pm 0.0039$ for the Bessell $R$-band. The correction factor for these parameters is 1.014 with errorbars of $\pm$0.006 from the $v\sin i$ of the brighter star, $\pm$0.007 from the $v\sin i$ of the fainter star, and $\pm$0.003 for the light ratio. Adding these uncertainties in quadrature gives a correction factor of $1.014 \pm 0.010$ and thus $K_{\rm A} = 158.9 \pm 2.3$\ms.

Our measurements of the physical properties of the WASP-2 system are given in Table\,\ref{tab:w02}. They were calculated from the parameters given in the previous section, plus $\FeH = 0.04 \pm 0.05$ from \citet{Me12mn}. We find results in good agreement with previous studies, the main difference being a slight increase in the measured mass of the planet and thus density and surface gravity.

\subsection{WASP-131} \label{sec:lc:w131}

\begin{figure} \includegraphics[width=\columnwidth,angle=0]{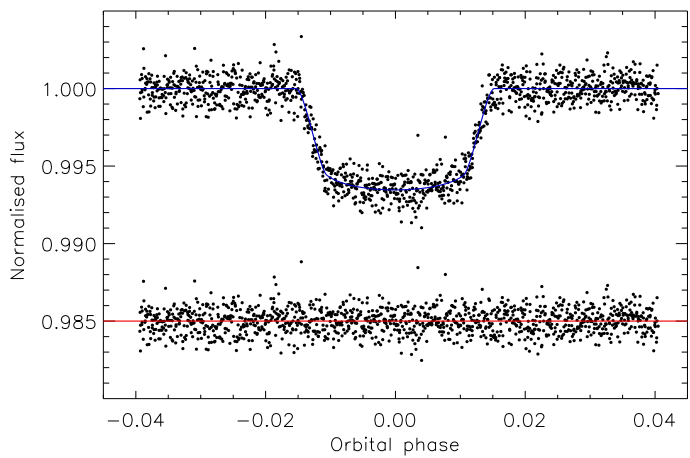}
\caption{\label{fig:lc:w131} Fit to the TESS light curve of WASP-131.
Only the fit with the planet orbiting star A is shown.} \end{figure}

The final system we have looked at in the current work is WASP-131. The companion star is relatively faint $\Delta K = 2.82 \pm 0.20$ but is very close (0.189\as) and thus was previously unknown. There is also a TESS light curve for this object that was not available to past analyses. The planetary nature of WASP-131 was discovered by \citet{Hellier+17mn} and the system is of interest because the planet has a very low density and surface gravity.

% Light ratio:      0.024735003     0.012363333
% Light contribution:      0.024137951     0.012064908

We obtained the TESS light curve and extracted the transits in the same way as in Section\,\ref{sec:lc:w20}. Our $\Delta K$ value corresponds to a light ratio between the two stars of $0.025 \pm 0.012$ and thus a third light of $L_3 = 0.024 \pm 0.012$.  This was used to model the TESS light curve for the scenario where the planet orbits the brighter star. We did not consider the planet-orbits-fainter star scenario because the analyses of WASP-8 and WASP-76 above make it clear that this possibility is not able to provide a good fit to the transit light curve when $\Delta K \goa 2.3$\,mag.

\begin{table} \centering
\caption{\label{tab:w131} Derived physical properties for the WASP-131 system. Where two sets of errorbars
are given, the first is the statistical uncertainty and the second is the systematic uncertainty.}
\begin{tabular}{lcc}
\hline
\hline
Parameter                 & \citet{Hellier+17mn}            & This work (planet              \\
                          &                                 & transits brighter star)        \\
\hline
$r_{\rm A}+r_{\rm b}$     &                                 & $0.1284 \pm 0.0049$            \\
$k$                       & $0.0815 \pm 0.0007$             & $0.08112 \pm 0.00083$          \\
$i$ ($^\circ$)            & $i = 85.0 \pm 0.3$              & $85.03 \pm 0.37$               \\
$r_{\rm A}$               &                                 & $0.1188 \pm 0.0045$            \\
$r_{\rm b}$               &                                 & $0.00964 \pm 0.00042$          \\
\hline
$M_{\rm A}$ (\Msun)       & $1.06 \pm 0.06$                 & $1.002 \pm 0.046 \pm 0.025$    \\
$R_{\rm A}$ (\Rsun)       & $1.53 \pm 0.05$                 & $1.526 \pm 0.064 \pm 0.013$    \\
$\log g_{\rm A}$ (cgs)    & $4.089 \pm 0.026$               & $4.072 \pm 0.033 \pm 0.004$    \\
$\rho_{\rm A}$ (\psun)    & $0.292 \pm 0.026$               & $0.282 \pm 0.032$              \\
$\tau$ (Gyr)              & 4.5 to 10                       & \err{7.2}{0.8}{1.6}{0.9}{1.0}  \\
\hline
$M_{\rm b}$ (\Mjup)       & $0.27 \pm 0.02$                 & $0.270 \pm 0.018 \pm 0.004$    \\
$R_{\rm b}$ (\Rjup)       & $1.22 \pm 0.05$                 & $1.204 \pm 0.056 \pm 0.010$    \\
$g_{\rm b}$ (\mss)        & $4.17 \pm 0.38$                 & $4.62 \pm 0.48$                \\
$\rho_{\rm b}$ (\pjup)    & $0.15 \pm 0.02$                 & $0.145 \pm 0.021 \pm 0.001$    \\
$T_{\rm eq}$ (K)          & $1460 \pm 30$                   & $1450 \pm 36$                  \\
$a$ (au)                  & $0.0607 \pm 0.0009$             & $0.0597 \pm 0.0009 \pm 0.0005$ \\
\hline
\end{tabular}
\end{table}

\subsubsection{Physical properties}

We obtained a correction factor for $K_{\rm A}$ under the assumption that all light from the fainter star contaminated the spectrum, and the measured $K_{\rm A}$ and $v\sin i$ are $30.5 \pm 1.7$\ms\ and $3.0 \pm 0.9$\kms, respectively \citep{Hellier+17mn}. We found a correction factor of 1.0269 with errorbars of $\pm$0.0001 from the $v\sin i$ of the brighter star, $\pm$0.0003 from the $v\sin i$ of the fainter star, and $\pm$0.0014 for the light ratio. The final uncertainty dominates all others so we adopted it as the errorbar for the correction factor. This yields a velocity amplitude of $31.3 \pm 1.8$\,m\,s$^{-1}$. This change in $K_{\rm A}$ is titchy and indicates that the contaminating light is sufficiently  small that it does not have a significant effect on the RVs.

We used the stellar atmospheric properties of $\Teff = 5950 \pm 100$\,K and $\FeH = -0.18 \pm 0.08$ from \citet{Hellier+17mn}, and our new values of $r_1$, $r_2$ and $i$ from the TESS light curve, to determine the physical properties of the system. Our results are shown in Table\,\ref{tab:w131} and are in excellent agreement with those from \citet{Hellier+17mn}. In particular, the uncertainties in most parameters are very similar between the two studies despite the availability of the TESS light curve, suggesting those in the previous study were underestimated.

%%%%%%%%%%%%%%%%%%%%%%%%%%%%%%%%%%%%%%%%%%%%%%%%%%%%%%%%%%%%%%%%%%%%%%%%%%%%%%%%%%%%%%%%%%%%%%%%%%%%%%%%%%%%%%%%%%%%%%%%%%%%%%%%%%%%%%%%%%%%%%%%%%%%%%

\section{Population studies of transiting planetary systems}

% \citet{Ciardi+15apj}:
% ignoring bias gives mean cor fac \langle X_R \rangle = 1.5
% good followup: \langle X_R \rangle = 1.2 for Kepler, expect 1.1 for brighter systems from K2 and TESS
% higher \langle X_R \rangle for more massive host stars (1.6 AFG, 1.2 KM)
% Driven to higher numbers by allowing the planet to orbit either star.
% Planet density more affected, by \langle X_R \rangle^3

% Ziegler+19aj: xr = 1.11 for planet orbits brighter and 2.55 for orbits fainter, avg 1.82 for equally likely either star

% Ziegler+18aj for Kepler: 1.09, 3.29, 2.18, and 1.54 overall. Trust only the latter number.

% Hirsch+17aj: xr=1.64
% Under the assumption that planets in bound binaries are equally likely to orbit the primary or secondary, we find a mean radius correction factor for planets in stellar multiples of X R = 1.65. If stellar multiplicity in the Kepler field is similar to the solar neighborhood, then nearly half of all Kepler planets may have radii underestimated by an average of 65%, unless vetted using high-resolution imaging or spectroscopy.
%
% Furlan+17aj:

\reff{The measured masses, radii and densities of the bulk population of transiting planets can be used to study their internal structure or formation mechanisms. It is important to correct for the effects of stellar multiplicity in such work, in order to decrease biases that might affect the results. Our SPHERE survey, coupled with the results presented in the current work, allowed us to investigate the size of these corrections.}

\reff{\citet{Ciardi+15apj} used stellar multiplicity rates for systems with separations of 1\as\ or less to infer that the mean planet radius correction factor, $\langle X_R \rangle$, is 1.5 for the \textit{Kepler} Objects of Interest (KOIs: planet candidates discovered using the \textit{Kepler} satellite). This quantity is given in the sense that one should multiply the measured radii of a population of planet candidates by $\langle X_R \rangle$ before comparing them to theoretical predictions of their physical properties. \citet{Ciardi+15apj} further broke this down into $\langle X_R \rangle \sim 1.6$ for hotter planet host stars (A-, F- and G-type) and $\langle X_R \rangle \sim 1.2$ for cooler ones (types K and M). They also noted that detailed follow-up observations including high-resolution imaging would bring $\langle X_R \rangle$ down from about 1.6 to 1.2, that brighter systems such as those discovered by K2 and TESS would have $\langle X_R \rangle \sim 1.1$, and that the planet density would be more strongly affected because $\rho_{\rm b} \propto R_{\rm b}^{~3}$.}

\reff{\citet{Furlan+17aj} discussed 1903 KOIs observed using new and published high-resolution images obtained with a variety of techniques, finding a total of 2297 nearby objects. They found that $\langle X_R \rangle$ was 1.09 or 3.09, under the assumption that the planets orbit the brighter or fainter stars, respectively. They expected the true value to be closer to the lower end of the range defined by these two values, and suggested $\langle X_R \rangle = 1.5$--2.0. \citet{Hirsch+17aj} concentrated on the stars within this sample hosting small planets ($R_2 < 5$\Rearth) and having companions within 2\as, and found $\langle X_R \rangle = 1.65$ under the assumption that each planet was equally likely to orbit either of the two stars.}

\reff{\citet{Ziegler+19aj} presented a study of 542 TESS transiting planet candidates using speckle imaging. They found $\langle X_R \rangle = 1.11$ if the planets orbited the brighter stars and $\langle X_R \rangle = 2.55$ if they orbited the fainter stars, with an average of $\langle X_R \rangle = 1.82$ if the planets were equally likely to orbit either star. A similar study of the KOIs by \citet{Ziegler+18aj} returned a value of $\langle X_R \rangle = 1.54$ for the last situation.}

\reff{With our survey of 45 transiting planetary systems (Paper\,I) and corrections to the measured physical properties (this work) we were in a position to calculate $\langle X_R \rangle$ for a sample of planetary systems. We restricted our analysis to hot Jupiters, which we defined in this case as planets of mass $>$0.2\Mjup\ and orbital period $<$12\,d. These restrictions removed four of the planets studied in Paper\,I (K2-24, K2-38, K2-39 and K2-99), leaving us with a sample of 41 objects.}

\reff{We first calculated $X_R$ for each of the six systems studied in this work, from the physical properties we measured with and without accounting for contamination. For WASP-20, where it is not yet clear which star hosts the planet, we calculated $X_R$ for both scenarios and took the average. For the remaining systems, where our light curve fits discount the possibility that the planet orbits the fainter star, we calculated $X_R$ under the assumption that the planet orbits the brighter star in each case. For the remaining 35 systems we adopted $X_R = 1.0$ as there are no detected companions bright enough to make a significant difference to the measured physical properties (see Section\,\ref{sec:lc:w131}). This means that our $\langle X_R \rangle$ is suitable for application to the bulk population of hot Jupiters.}

\reff{We find a mean planet radius correction factor of $\langle X_R \rangle = 1.009 \pm 0.045$ (standard deviation $\sigma = 0.016$). The uncertainty has been propagated from the individual values for each $X_R$, and does not account for the small sample size. This is much smaller than found in previous works (see above). The first reason for this is that we included all observed systems rather than just those with a known companion, thus making our results more widely applicable to populations of planetary systems. The second reason is that large values of $X_R$ are obtained when a planet orbits the fainter of two stars, but we were able to rule out this possibility in all cases but one. Most previous works have assumed an equal probability for which star hosts the planet, which ignores situations when the data are only consistent with one scenario, and also neglects changes in planet occurrence rates as a function of host star mass.}

\reff{We were also in a position to calculate corrections to planet mass. Using the same procedure as above, and accounting for the fact that some planet masses are unchanged because the light from the nearby star fell outside the entrance aperture of the spectrograph, we find $\langle X_M \rangle = 1.031 \pm 0.019$ ($\sigma = 0.019$). This is once again a small correction, but is driven to a larger value than $\langle X_R \rangle$ because of the large effect of contaminating light in the case of WASP-20\,b (Table\,\ref{tab:w20}).}

\reff{Finally, we have calculated the mean planet density correction factor to be $\langle X_\rho \rangle = 0.995 \pm 0.046$ ($\sigma = 0.046$). This result is counter to the expectation that $\langle X_\rho \rangle$ would be significantly larger than $\langle X_R \rangle$, and this occurs for two reasons. First, contaminating light causes both the measured planet mass and radius to decrease, and partial cancellation of these effects yields values of $X_\rho$ that are not much larger than unity (i.e.\ contamination causes the measured density to decrease). Second, in some cases (e.g.\ WASP-70 and WASP-8), the source of contaminating light is several arseconds from the planetary system. In such cases it affects the light curve but not the RVs, resulting in values of $X_\rho$ that are below unity (i.e.\ contamination causes the measured density to increase). The net result is that $\langle X_\rho \rangle$ is approximately unity for the sample considered here, but with a large scatter.}

\reff{We have therefore obtained mean correction factors that can be applied to the masses, radii, and densities of transiting hot Jupiter systems. Our sample selection (Paper\,I) was based only on target brightness and observability: it was agnostic about the physical properties of the planetary systems or whether a nearby companion was already known. Our sample is therefore representative of bright hot Jupiters predominantly discovered using ground-based surveys. The mean correction factors we have derived are suitable for application to similar samples of transiting planets, but not to samples with significantly different properties. In particular, the mean correction factors are likely to be larger for smaller planets because their shallower transits can be adequately fitted with a wider range of contamination levels, smaller for more nearby planetary systems because a larger fraction of bound companions will be resolved, and more scattered for planetary systems in crowded areas of the sky due to the wider variety of systemic velocities of contaminating objects.}

%%%%%%%%%%%%%%%%%%%%%%%%%%%%%%%%%%%%%%%%%%%%%%%%%%%%%%%%%%%%%%%%%%%%%%%%%%%%%%%%%%%%%%%%%%%%%%%%%%%%%%%%%%%%%%%%%%%%%%%%%%%%%%%%%%%%%%%%%%%%%%%%%%%%%%

\section{Summary and conclusions} \label{sec:conc}

We have presented a detailed analysis of six transiting planetary systems in order to account for the effect of fainter nearby stars on the measured physical properties of the system. For one of these systems the nearby star was discovered in Paper\,I, and for the remaining five it was detected in previous studies. Contaminating light affects the photometric properties of a system: it dilutes the transit depth and biases the measured planet radius to lower values. It also affects spectroscopic analysis by contaminating the CCFs from which RVs are measured, causing a decrease in the RV variation and thus an underestimate of the planet mass. We used an existing approach to ameliorate the photometric bias and presented a new method to account for the RV bias.

WASP-20 is the system most affected because its nearby star is relatively bright. Our analysis of this system agrees well with that of \cite{Evans++16apj} and is an improvement because of the availability of a high-quality light curve from the TESS satellite. The second system, WASP-70, has a contaminant that is sufficiently distant to leave the spectroscopy of this system unaffected, and sufficiently faint to have only a small effect on the photometric properties of the planetary system. A similar story occurs for WASP-8, although a significant improvement in its characterisation is achieved using the TESS light curve. The final three systems, WASP-76, WASP-2 and WASP-131, all have contaminating stars within 0.7\as\ that are nevertheless sufficiently faint to make little difference to measurements of their physical properties. The updated physical properties of the systems have been lodged in the TEPCat catalogue\footnote{\texttt{http://www.astro.keele.ac.uk/jkt/tepcat/}} \citep{Me11mn}.

Looking at Paper\,I, we see no other systems that would significantly benefit from the analysis of individual objects as presented above. Our results for WASP-131 showed that a contaminating star fainter by $\Delta K = 2.80$\,mag is too faint to make much difference to the measured physical properties of planetary systems such as those studied in the current work. Only one more object in Paper\,I has a $\Delta K$ smaller than this: HAT-P-41 has $\Delta K = 2.50 \pm 0.21$\,mag; and this star was already known and accounted for \citep{Hartman+12aj}.

\reff{We have used our sample of 45 transiting hot Jupiter systems, and the corrections to their measured properties needed to account for contaminating light, to determine mean correction factors for samples of planet masses, radii, and densities. We find $\langle X_M \rangle = 1.031 \pm 0.019$, $\langle X_R \rangle = 1.009 \pm 0.045$ and $\langle X_\rho \rangle = 0.995 \pm 0.046$, respectively. The radius correction is much smaller than found by other studies, primarily because we were able to reject the possibility that the planet orbits the fainter star for five out of the six systems we studied in detail. The mass and density corrections are also small, and to our knowledge are the first ones to be published. The mean correction factors will depend on the population of objects under consideration, specifically on the planet radius, system distance, and sky position (via the amount of field star contamination).}

We conclude that it is important to obtain high-resolution observations of transiting planetary systems in order to detect cases like WASP-20 and Kepler-14, where the physical properties are strongly affected by the presence of a nearby star, but that these cases are sufficiently rare that they will have a negligible influence on studies of the overall population of planetary systems. This is good news.

%%%%%%%%%%%%%%%%%%%%%%%%%%%%%%%%%%%%%%%%%%%%%%%%%%%%%%%%%%%%%%%%%%%%%%%%%%%%%%%%%%%%%%%%%%%%%%%%%%%%%%%%%%%%%%%%%%%%%%%%%%%%%%%%%%%%%%%%%%%%%%%%%%%%%%

\begin{acknowledgements}
The  research of AJB leading to these results has received funding from the European Research Council under ERC Starting Grant agreement 678194 (FALCONER). The following internet-based resources were used in research for this paper: the ESO Digitized Sky Survey; the NASA Astrophysics Data System; the SIMBAD database operated at CDS, Strasbourg, France; and the ar$\chi$iv scientific paper preprint service operated by Cornell University.
\end{acknowledgements}

%%%%%%%%%%%%%%%%%%%%%%%%%%%%%%%%%%%%%%%%%%%%%%%%%%%%%%%%%%%%%%%%%%%%%%%%%%%%%%%%%%%%%%%%%%%%%%%%%%%%%%%%%%%%%%%%%%%%%%%%%%%%%%%%%%%%%%%%%%%%%%%%%%%%%%

\bibliographystyle{aa}
% \bibliography{jkt}

%%%%%%%%%%%%%%%%%%%%%%%%%%%%%%%%%%%%%%%%%%%%%%%%%%%%%%%%%%%%%%%%%%%%%%%%%%%%%%%%%%%%%%%%%%%%%%%%%%%%%%%%%%%%%%%%%%%%%%%%%%%%%%%%%%%%%%%%%%%%%%%%%%%%%%
\end{document}